\documentclass[aps,showpacs,nofootinbib,tightenlines,12pt]{revtex4}
\usepackage{amsmath}
\usepackage{amssymb}
\usepackage{epsfig}
\usepackage{graphicx}
\begin{document}
\def \ctp{  Commun. Theor. Phys. }
\def \epjc{  Eur. Phys. J. C }
\def \jpg{  J. Phys. G }
\def \npb{  Nucl. Phys. B }
\def \plb{  Phys. Lett. B }
\def \prd{  Phys. Rev. D }
\def \prl{  Phys. Rev. Lett.  }
\def \pr{   Phys. Rep. }
\def \rmp{  Rev. Mod. Phys. }
\def \ptp{  Prog. Theor. Phys. }
\def \zpc{  Z. Phys. C }
\newcommand{\acp}{{\cal A}_{CP}}
\newcommand{\be}{\begin{equation}}
\newcommand{\ee}{\end{equation}}
\newcommand{\beq}{\begin{eqnarray}}
\newcommand{\eeq}{\end{eqnarray}}
\newcommand{\non}{\nonumber\\ }

\title{CP violation of the two-body charmless hadronic $B$ decays
          in the minimal supergravity model}
\author{Wenjuan Zou}
\email{zouwenjuan@email.njnu.edu.cn}
\author{ Zhenjun Xiao}
\email{xiaozhenjun@njnu.edu.cn} \affiliation{Department of Physics
and Institute of Theoretical Physics, Nanjing Normal University,
Nanjing, Jiangsu 210097, P.R.China}
\date{\today}
\begin{abstract}
By choosing two typical input parameter points in the minimal
supergravity (mSUGRA) model and using the QCD factorization (QCDF)
approach, we studied the supersymmetric effects to the CP
violation of the two-body charmless hadronic $B$ meson decays. We
found that though the SUSY contributions can give large
corrections to the CP asymmetries for some decay channels, they
could not be distinguished experimentally from the SM values
because of the large theoretical errors dominated by calculating
the annihilation contributions in the QCDF approach.
\end{abstract}


\maketitle

As one of the main goals of $B$ experiments,
measurements of CP asymmetries in various $B$ decay processes are
at the center of attention. Since in the standard model (SM) all
the measured CP asymmetries have to be consistently explained by
the complex phase $\delta$ in the Cabibbo-Kobayashi-Maskawa (CKM)
mixing matrix, precise measurements of CP asymmetries at the
current $B$ factories can be used to test the CKM mechanism and
sequentially the SM\cite{cpvb,qcdf}.

Experimentally, Belle collaboration has given the first
experimental evidence for the direct CP violation in the decay
mode $B\to \pi^+\pi^-$\cite{kabe04}. And presently large direct CP
violation in neutral B meson decay $B^0 \to K^+\pi^-$  has been
measured by both BaBar\cite{babar} and Belle\cite{ycbe04}
collaborations, but no CP violation has been found for the charged
$B$ meson decay $B^+ \to K^+ \pi^0$. The world averages given by
the Heavy Flavor Averaging Group (HFAG) \cite{hfag05} are
\beq
{A}_{CP} (B^0  \to K^+ \pi ^ -  ) &=& -0.115 \pm 0.018,
\label{eq:kpi1} \non { A}_{CP} (B^+  \to K^+  \pi ^ 0  ) &=& -0.04
\pm 0.04 \label{eq:kpi0}.
\eeq
In the SM and with the QCD factorization (QCDF)  approach, the CP-violating
asymmetry for
these two channels should be naturally very similar in size and
positive in sign for a set of favored hadronic parameters
\cite{qcdf}. In the perturbative QCD (PQCD)  factorization
approach\cite{pqcd}, however, the SM predictions are
\cite{sanda01} $A_{CP} (B^0  \to K^+ \pi ^ - )\approx -0.18$ and
$A_{CP} (B^0  \to K^+ \pi ^ - )\approx -0.15$: the first
prediction agrees well with the measured value as given in
Eq.~(\ref{eq:kpi1}), but the second one shows a large deviation
with the data.

For the measurement of the time-dependent CP asymmetries,  great
efforts have been made by BaBar, Belle and other collaborations.
The recently reported measurements of time-dependent CP
asymmetries in $B\to \phi K_s$ decays and other s-penguin modes
lead to \cite{hfag05}
\beq sin(2\beta)_{ave}=0.50\pm0.06
\label{phi}, \label{eq:beta-s}
\eeq
which is a $2.6\sigma$
deviation from  that for $B\to J/\psi K_s$ and other charmonium
modes,
\beq sin(2\beta)_{ave}=0.687\pm0.032~. \label{psi}
\label{eq:beta-c}
\eeq
Therefore, something must be there to
account for the large deviations. Since the $B\to \phi K_s$ and
other s-penguin modes are induced at the one loop level, while
$B\to J/\psi K_s$ and other charmonium modes are governed by tree
level decays, it is tempting to expect that the  new physics (NP)
contributions are more significant to the former and do not show
up clearly to the latter.

Recently, many works about various NP models have been done in the
$B$ meson system \cite{jcx04}. In our  previous papers
\cite{zw04}, we have studied  the minimal supergravity (mSUGRA)
model by calculating the branching ratios of $B \to M_1 M_2 $
($M_i$ stands for the light pseudo-scalar (P) or vector (V)
mesons) decays, and have found some interesting results. Based on
the previous works \cite{zw04}, here we will try to calculate how
much can the SUSY contributions in the mSUGRA model affect the CP
asymmetries of the $B \to M_1 M_2$ decays by using the same input
parameters as listed in the Appendix  of Ref.~\cite{zw04}.

In the mSUGRA model\cite{msugra}, only four continuous free
parameters and an unknown sign are left, say $\tan\beta$,
$m_{\frac{1}{2}}$, $m_{0}$, $A_{0}$ and $sign(\mu)$. The new
effects   on the rare $B$ meson decays from this model will
manifest themselves through the corrections to the Wilson
coefficients of the same operators involved in the SM calculation.
In Ref.~\cite{zw04}, we have found that the SUSY contributions to
the Wilson coefficients $C_k (k=3\sim6)$ are always small and can
be neglected safely. But for the Wilson coefficients
$C_{7\gamma}(m_b)$ and $C_{8g}(m_b)$, the SUSY contributions can
be rather large and even make them have an opposite sign with the
SM ones.

Similar to Ref.~\cite{zw04}, considering the constraints
coming from the well measured $B\to X_s \gamma$ decays and so on,
we take two typical parameter points, say Case A and B as listed
in the following tabular. For Case C in Ref.~\cite{zw04}, the
Wilson coefficient $C_{7\gamma}(m_b)$ in the mSUGRA model is also
SM-like (negative) as that in Case A. Hence in our calculations we
will only consider Case A and B. The numerical values of the ratio
$R_7=C_{7\gamma}(m_b)/C_{7\gamma}^{SM}(m_b)$ for the two cases
have also been given in the tabular. It is easy to see that the
Wilson coefficient $C_{7\gamma}(m_b)$ in the mSUGRA model is
SM-like (negative) for Case-A, but nonstandard (positive) for
Case-B.

\begin{center}
\begin{tabular}{c|cccccccccc|ccccc} \hline  \hline
\multicolumn{1}{c|}{Case}& \multicolumn{2}{c}{$m_0$\qquad
}&\multicolumn{2}{c}{$m_{\frac{1}{2}}$\qquad }&
\multicolumn{2}{c}{$A_0$\qquad }&\multicolumn{2}{c}{$\tan\beta$}&
\multicolumn{2}{c|}{$Sign[\mu]$}&
\multicolumn{5}{c}{$ R_7$} \\
\hline A&\multicolumn{2}{c}{$300$}&\multicolumn{2}{c}{$300$}&
\multicolumn{2}{c}{$0$}&\multicolumn{2}{c}{$2$}&\multicolumn{2}{c|}{$-$}&
\multicolumn{5}{c}{$1.10$} \\
B&\multicolumn{2}{c}{$369$}&\multicolumn{2}{c}{$150$}&
\multicolumn{2}{c}{$-400$}&\multicolumn{2}{c}{$40$}&\multicolumn{2}{c|}{$+$}&
\multicolumn{5}{c}{$-0.93$} \\ \hline \hline
\end{tabular}
\end{center}

Just take a look at the two typical sets of SUSY parameters, both
of them are chosen to be real for the constraint condition coming
from the measured electric dipole moment (EDM) of neutron and
electron. Therefore no new weak phase has been introduced and the
mechanism of CP violation in the mSUGRA  model is the same as that
in the SM. But when we calculate the amplitudes of $B$ decays by
using the QCDF approach \cite{qcdf}, the real and imaginary part
of the factorized coefficient $a_i(M_1M_2)$ or $\alpha_i(M_1M_2)$
(See Ref.~\cite{qcdf} for detailed definitions) will be modified
by the SUSY contributions and hence the CP-violating parameter
will be affected in this model.

We begin with a discuss of the CP asymmetries in $B\to PP$ decays.
Firstly, for those channels which have only direct CP violations,
one can define
\begin{equation}
{\cal  A}_{CP} = \frac{\Gamma(\bar{B} \to \bar{f}) -\Gamma(B \to
 f)}{ \Gamma(\bar{B} \to \bar{f}) + \Gamma(B \to
f)}.\label{eq:acpp}
\end{equation}

As we have discussed in Ref.~\cite{zw04}, the potential SUSY
contributions are mainly embodied in $\alpha_4^p(M_1M_2)$ and
$\alpha_{4,ew}^p(M_1M_2)$. Therefore, we naturally expect a
 large new physics corrections to those penguin
dominated $B$ meson decays. According to our calculations, things
are absolutely so. For the tree-dominated decay channels, say
$B^\pm\to \pi^\pm(\pi^0,\eta^{(')})$, the SUSY corrections are
less than $10\%$ for both Case A and B. However for other channels
which are penguin-dominated, we can see from Table \ref{bupp}: (a)
in Case A, all the corrections are still small and less than
$10\%$ but in Case B the corrections are large and the largest one
is about $-35\%$ for $B\to \pi^\pm K^\mp$. (b)Though the
corrections in Case B are large for those channels, experimentally
they can not be distinguished from the SM values for they are all
of the same order of magnitude as SM values. (c) For the very
interesting decay $B\to \pi^\pm K^\mp$,  $A_{CP}$ in both  the SM
and mSUGRA model  have a different sign with the data in
Eq.~(\ref{eq:kpi1}). If we consider the large uncertainties from
the annihilation contributions, $A_{CP}$ is
\begin{equation}
A_{CP}(\pi^\pm K^\mp) =\left\{
\begin{array}{ll}
{4.5 ^{+8.9}_{-9.4},
\ \  {\rm{ \ \ SM}}} \\
{4.9 ^{+9.5}_{-10.2},
 \ \  {\rm{   Case-A}}} \\
{3.0^{+5.8}_{-6.1}.
   \ \  {\rm{ \ \  Case-B}}} \\
\end{array}
\right.\\
\end{equation}
and can only be consistent with the data within about three
standard deviations. However, in Refs.~\cite{hca05}, by using the
PQCD approach or considering the final state interaction (FSI)
modifications to short-distance (SD) predictions in QCDF approach,
the authors found the consistency between the theory and the
experiments for $B\to \pi^\pm K^\mp$ decay. For more details, one
can see these references.

\begin{table}[h]   
\doublerulesep 1.5pt \caption{Theoretical results in percent for
those penguin dominated and having only direct CP violations
$A_{CP}$ channels of $B\to PP$ in the SM and mSUGRA model, where
superscripts ``f+a" and ``f" denote the predictions with or
without weak annihilation contributions considered.} \label{bupp}
\begin{center}
\begin{tabular} {l|cc|c|c} \hline  \hline
\ \ \ \ Decay &
\multicolumn{2}{|c|}{SM }& Case A & Case B \\
\cline{2-5} \ \ \ \ Modes &$\mathcal{A}_{CP}^{f}$
&$\mathcal{A}_{CP}^{f+a}$
&$\mathcal{A}_{CP}^{f}$&$\mathcal{A}_{CP}^{f}$
\\ \hline \hline
$B_{d}\to \pi^{\pm}K^{\mp}$&5.36&4.54&5.75 &3.47 \\
 $B^{\pm}\to \pi^{0}K^{\pm}$&8.28&7.45&8.70&6.09\\
$B^{\pm}\to \pi^{\pm}K^{0}$&1.00 &0.91&1.03&0.81\\
 $B^{\pm}\to K^{\pm}\eta$&$-13.7$&$-12.8$&$-14.1$&$-11.5$\\
 $B^{\pm}\to K^{\pm}\eta^{'}$&2.38&2.04&2.46 &1.94 \\
$B^{\pm}\to K^{\pm}K^{0}$&$-26.8$&$-24.0$&$-27.6$&$-22.0$\\
 \hline\hline
\end{tabular}
\end{center}
\end{table}

Secondly, for the CP asymmetries of the left $B\to PP$ channels,
they are time-dependent and can be described by
\begin{eqnarray}
{\cal  A}_{CP}(t) &=& \frac{\Gamma(\bar{B}^0(t) \to \bar{f} )
 -\Gamma(B^0(t) \to f )}{
\Gamma(\bar{B}^0(t) \to \bar{f}) + \Gamma(B^0(t) \to f)} \non &=&
S_f \sin(\Delta M t)-C_f \cos(\Delta Mt) \label{eq:acp0}
\end{eqnarray}
where $C_f$ and $S_f$ represent the direct and the mixing CP
asymmetry, respectively, and they are given by
\begin{equation}
C_f= \frac{1-|\lambda_{CP}|^2 }{1 + |\lambda_{CP}|^2}, \ \
S_f=\frac{2 Im(\lambda_{CP})}{1 +
|\lambda_{CP}|^2}. \label{eq:aep}\\
\end{equation}
The parameter $\lambda_{CP}$ is defined by
\begin{equation}\label{lam}
\lambda_{CP}=\frac{V^*_{tb}V_{td}}{V_{tb}V^*_{td}} \frac{ {\cal
A}(\bar{B}^0(0)\to \bar{f} )}{ {\cal  A}(B^0(0)\to f )}.
\end{equation}

Through the numerical calculations, we found that for those
neutral $B_d \to PP$ decays,  the SUSY contributions to the two CP
asymmetric parameters $C_f$ and $S_f$ are generally small or
moderate in both Case A and  B. The largest correction is only
about $26\%$ to $C_f$ of $B_d\to K_S^{0}\pi^{0}$ and make it vary
from $3.8\%$ to $2.8\%$. Obviously, so small correction can not be
distinguished experimentally and can be masked easily by the
uncertainty coming from the annihilation contributions .

\begin{table}[h]
\doublerulesep 1.5pt \caption{The same as Table \ref{bupp}, but
for some $B\to PV$ channels.}\label{bupv}
\begin{center}
\begin{tabular} {l|cc|c|c} \hline  \hline
\ \ \ \ Decay &
\multicolumn{2}{|c|}{SM }& Case A & Case B \\
\cline{2-5} \ \ \ \ Modes &$\mathcal{A}_{CP}^{f}$
&$\mathcal{A}_{CP}^{f+a}$
&$\mathcal{A}_{CP}^{f}$&$\mathcal{A}_{CP}^{f}$
\\ \hline \hline
$B^{\pm}\to \pi^{\pm}K^{*0}$&2.01&1.71&2.19&1.27\\
 $B^{\pm}\to \pi^{0}K^{*\pm}$&9.26  &8.56 &9.81 &6.32\\
$B^{\pm}\to K^{*\pm}\eta$&3.19&2.98 &3.21&3.08\\
$B^{\pm}\to  K^{*\pm}\eta^{'}$&$-27.5$&$-19.9$&$-25.1$&$-64.5$\\
$B^{\pm}\to K^{\pm}\rho^{0}$&$-13.5$&$-11.6$&$-13.1$&$-17.2$\\
$B^{\pm}\to  K^{0}\rho^{\pm}$&0.12&0.16&0.12&0.19\\
$B^{\pm}\to  K^{\pm}\omega$&$-6.61$&$-5.90$&$-6.43$&$-8.21$\\
$B^{\pm}\to K^{\pm}\phi$&2.12&1.75&2.27&1.44\\
$B^{\pm}\to K^{\pm}K^{*0}$&$-52.7$&$-46.0$&$-56.7$&$-34.4$\\
$B^{\pm}\to K^{0}K^{*\pm}$&$-2.55$&$-2.92$&$-2.43$&$-4.03$\\
$B_{d}\to \pi^{\pm}K^{*\mp}$&$-0.32$&$-0.37$&$-0.28$&$-0.26$\\
$B_{d}\to \pi^{0}\bar{K}^{*0}$&$-15.4$&$-12.9$&$-18.1$&$-7.34$\\
$B_{d}\to K^{\pm}\rho^{\mp}$&$-4.51$&$-3.25$&$-4.20$&$-8.18$\\
$B_{d}\to \bar{K}^{*0}\eta$&4.93&4.45&4.97&4.69\\
$B_{d}\to \bar{K}^{*0}\eta^{'}$&$-11.2$&$-8.15$&$-10.4$
&$-23.8$\\
 \hline\hline
\end{tabular}
\end{center}
\end{table}

\begin{table}[h]
\doublerulesep 1.5pt \caption{The same as Table \ref{bupp}, but
for CP asymmetries parameter $C_f$ and $S_f$ of some $B_{d}\to PV$
channels.}\label{bdpv}
\begin{center}
\begin{tabular} {l|c|cc|c|c} \hline  \hline
 Decay Modes &\ \ &
\multicolumn{2}{|c|}{SM }& Case A & Case B \\
\cline{1-6}
$B_{d}\to \pi^{0}\rho^{0}$&$C_f$&$-4.47$ &$-3.03$&$-2.94$&$-15.8$\\
\ \ &$S_f$&$-32.1$ &$-35.8$&$-35.2$&$-8.92$\\\cline{1-6}
$B_{d}\to \pi^{0}\omega$&$C_f$&74.4 &$94.6$&$73.0$&$61.6$ \\
\ \ &$S_f$& $-62.5$&32.3&$-56.8$&$-64.9$\\\cline{1-6}
$B_{d}\to \eta\rho^{0}$&$C_f$&$6.28$&$1.94$&$-4.51$&$47.6$ \\
\ \ &$S_f$&$-29.3$ &$-32.7$&$-52.4$&87.9\\\cline{1-6}
$B_{d}\to \eta^{'}\rho^{0}$&$C_f$&$46.5$&$37.3$&$44.8$&$58.1$\\
\ \ &$S_f$&$-69.5$ &$-63.3$&$-73.6$&$-24.6$\\\cline{1-6}
$B_{d}\to \eta\omega$&$C_f$&$8.36$ &$5.48$&$7.14$&$16.9$ \\
\ \ &$S_f$&$-41.5$ &$-38.0$&$-39.4$&$-55.6$\\\cline{1-6}
$B_{d}\to \eta^{'}\omega$&$C_f$&$-18.4$&$-17.9$&$-19.3$&$-11.5$ \\
\ \ &$S_f$&$-28.9$ &$-25.5$&$-27.0$&$-42.9$\\\cline{1-6}
$B_{d}\to K_{s}^{0}\rho^{0}$&$C_f$&$-9.15$&$-7.91$&$-8.85$&$-12.3$ \\
\ \ &$S_f$&$-62.1$ &$-63.8$&$-62.6$&$-57.1$\\\cline{1-6}
$B_{d}\to K_{s}^{0}\omega$&$C_f$&$9.71$ &$8.26$&$9.36$&$13.5$ \\
\ \ &$S_f$&$-89.8$ &$-87.9$&$-89.5$&$-93.9$\\\cline{1-6}
$B_{d}\to K_{s}^{0}\phi$&$C_f$&$-2.12$ &$-1.85$&$-2.27$&$-1.44$ \\
\ \ &$S_f$&76.9 &76.6&76.9&76.9\\\cline{1-6}
 \hline\hline
\end{tabular}
\end{center}
\end{table}
We now take a look at $B\to PV$ mode. Different from $B\to PP$
mode, the SUSY correction to this mode can interfere with the SM
counterparts constructively or destructively \cite{zw04}.
Through the numerical results, we found  the new physics
corrections to the CP violations in Case B are large for most
channels. For those channels having only direct CP violation, the
penguin dominated ones are affected a lot. As shown in Table
\ref{bupv}, in Case B the largest SUSY corrections is about
$135\%$ for  $B^{\pm}\to K^{*\pm}\eta^{'}$ channel and make the
size of its CP violation increased from $28\%$ to $65\%$. Such
large corrections may be measured experimentally.

For other neutral $B_d \to PV$ decays, they have both direct CP
violation $C_f$ and mixing CP asymmetry $S_f$. For the
CP-violating parameters of $B\to (\pi,\eta^{(')})\phi$ decays,
since these channels receive no contributions from electromagnetic
or chromo-magnetic penguin operators where the SUSY contributions
are entered, they remain almost unchanged in the mSUGRA model. For
other channels as shown in Table \ref{bdpv}, the direct CP
violations of most channels in Case B are greatly affected by the
SUSY corrections. The largest corrections even reach a factor of 7
for the decay $B^{0}\to \eta\rho^{0}$, about $253\%$ increase for
$B^{0}\to \pi^{0}\rho^{0}$ and $100\%$ enhancement for $B^{0}\to
\eta\varpi$. As to the indirect CP violation $S_f$ in Case B,  the
SUSY corrections are also large. For $B_d\to \eta \rho^{0}$ decay,
for example, the sign of its indirect CP violation has been
changed and the size also increased by a factor of $3$. But for
the very interesting channels $B\to \phi K_s$, the SUSY
contributions make little effects and hence the ``$\phi K_s$''
puzzle as mentioned above still cannot be solved here.
\begin{table}[t]
\doublerulesep 1.5pt \caption{The same as  Table \ref{bupp}, but
for some $B\to VV$ channels.}\label{bdvv}
\begin{center}
\begin{tabular} {l|c|c|c} \hline  \hline
 \multicolumn{1}{c|}{$A^{dir}(B\to VV)$}
& \multicolumn{1}{|c|}{ SM } &Case A& Case B
  \\ \hline\hline
$A_{CP}(\bar{B}^0\to K^{*-}\rho^+)$&17.8&19.3&10.5
 \\
$A_{CP}(\bar{B}^0\to \bar{K}^{*0}\rho^0)$&$-21.1$&$-23.4$
&$-11.9$ \\
$A_{CP}(B^-\to K^{*-}\rho^0)$&18.7&19.6 &13.8
 \\
$A_{CP}(B^-\to\bar{K}^{*0}\rho^-)$&1.57&1.63 &1.25
\\
\hline $C_{f}(\bar{B}^0\to \bar{K}^{*0}K^{*0})$&$-28.2$
&$-29.4$ &$-21.6$ \\
$A_{CP}(B^-\to K^{*-}K^{*0})$&$28.2$&$29.4$ &$21.6$
 \\
\hline $A_{CP}(\bar{B}^0\to \bar{K}^{*0}\omega)$&12.8 &13.4
&9.68  \\
$A_{CP}(B^-\to K^{*-}\omega)$&32.5&34.3 &22.4
\\
\hline $A_{CP}(\bar{B}^0\to \bar{K}^{*0}\phi)$&1.75&1.81&1.38
 \\
$A_{CP}(B^-\to K^{*-}\phi)$&1.75&1.81&1.38\\ \hline \hline
\end{tabular}
\end{center}
\end{table}

At last, let us talk about $B\to VV$ mode in brief. For $B\to VV$,
we first give two remarks: (a)from Eq.(\ref{lam}) one can see that
the parameter $\lambda_{CP}$ for $B\to VV$ mode is
helicity-dependent since three decay amplitudes with
$\lambda=(0,\pm1)$ are different. It follows that theoretically
the indirect CP violation $S_f$ can  be given only when the two
vector mesons in the final state have a certain helicity. Hence in
our calculation we only give the direct CP violation ($A_{CP}$ or
$C_f$) of $B\to VV$ in Table \ref{bdvv}. (b)the annihilation
amplitude in the VV case does not gain a chiral enhancement of
order $M_P^2/(m_qm_b)$ as that in $B\to PP$ and $PV$ modes.
Therefore, it is truly power suppressed in heavy quark limit and
we have ignored such contributions in our calculation.

Similar to $B\to PP$, only in the QCD penguin-dominated channels
can their CP violation be affected a lot. From Table \ref{bdvv},
in Case B the mSUGRA predictions of CP violations for all the
listed channels become smaller than corresponding SM predictions.
As to $\bar{B}^0\to \bar{K}^{*0}\rho^0$, the SUSY contributions in
Case B can even make its CP-violating parameter decreased by about
$44\%$. However, these SUSY contributions are of the same order of
magnitude as SM values and therefore almost impossible to be distinguished
experimentally.

To conclude, we have computed the CP asymmetries for two-body
charmless $B \to M_1M_2$ decays in the  mSUGRA model based on our
previous works\cite{zw04}. Since the SUSY phases in the mSUGRA
model are so small as to be ignored safely and the Yukawa
couplings are the main source of the flavor structure, we found
that the SUSY contributions to the CP asymmetries in charmless $B$
decays are generally small except in most $B \to PV$ channels.
Currently, people have tried to measure the CP violation in $B$
decays, but the data are always error-weighted and consistent with
zero except for $B\to \pi^\pm K^\mp$. Moreover, large theoretical
uncertainties also exist for example in the QCDF approach per se where
the weak annihilations and other potential power corrections are
not well calculated. Therefore, low experimental statistics and
large theoretical uncertainties together prevent us from testing
the mSUGRA model through studies of the CP-violating asymmetries
at present. We are waiting for great progress in both the theory
and experiment.


\end{document}